\documentclass{aa}  

\usepackage{graphicx}
\usepackage{txfonts}
\usepackage[colorlinks=true,     linkcolor=blue, citecolor=blue, filecolor=blue, urlcolor=blue]{hyperref}
\linenumbers

\begin{document}

   \title{Dynamical confirmation of a black hole in the X-ray transient Swift J1727.8$-$1613}
   \titlerunning{Dynamical confirmation of a black hole in the X-ray transient Swift J1727.8$-$1613}
   \author{D. Mata S\'anchez\inst{1,2}
          \and
          M. A. P. Torres\inst{1,2}
          \and
          J. Casares\inst{1,2}
          \and
          T. Mu\~noz-Darias\inst{1,2}          
          \and
          M. Armas Padilla\inst{1,2}
          \and
          I. V. Yanes-Rizo\inst{1,2}
          }

   \institute{Instituto de Astrof\'isica de Canarias, E-38205 La Laguna, Tenerife, Spain\\
              \email{matasanchez.astronomy@gmail.com, dmata@iac.es}
         \and
             Departamento de Astrof\'isica, Univ. de La Laguna, E-38206 La Laguna, Tenerife, Spain\\
             }

   \date{Received August 22, 2024; accepted November 12, 2024}

  \abstract
  {The X-ray transient Swift~J1727.8$-$1613 ended its 10-month discovery outburst in June of 2024, when it reached an optical brightness comparable to pre-discovery magnitudes. With the aim of performing a dynamical study, we launched an optical spectroscopy campaign with the GTC telescope. We detected the companion star and constructed its radial velocity curve, yielding a binary orbital period of $P_{\rm orb}= 10.8038\pm 0.0010\, {\rm h}$ and a radial velocity semi-amplitude of $K_{2}=390\pm 4\,{\rm km\, s^{-1}}$. This results in a mass function of $f(M_{\rm 1})=2.77\pm 0.09\, {\rm M_{\rm \odot}}$. Combined with constraints on the binary inclination, it sets a lower limit on the compact object mass of $M_{\rm 1}>3.12\pm 0.10\, {\rm M_{\rm \odot}}$, dynamically confirming the black hole nature of the accretor. A comparison of the average spectrum in the rest frame of the companion with synthetic stellar templates supports a K4V donor that is partially ($74\%$) veiled by the accretion disc. A refined distance measurement of $3.4\pm 0.3\, {\rm kpc}$, together with the astrometric proper motion and the systemic velocity derived from the radial velocity curve ($ \gamma= -181 \pm 4 \,\rm{km\, s^{-1}}$), supports a natal kick velocity of $v_{\rm kick}=210^{+40}_{-50}\,\rm{ km\,s^{-1}}$, at the upper end of the observed distribution.}

   \maketitle
%

\section{Introduction}

Black hole transients (BHTs) are traditionally discovered when they enter into outburst, epochs when their accretion discs become several magnitudes brighter across the electromagnetic spectrum. Early analysis of their spectral and photometric properties enables the preliminary classification of these objects as low-mass X-ray binaries (LMXBs): systems formed by a low-mass, Roche-lobe-filling companion star transferring mass onto a compact object (whether a black hole (BH) or a neutron star (NS)). Different techniques have been developed to pinpoint the nature of the accretor, but dynamical studies, which are only feasible during the fainter quiescence, set the gold standard (see e.g. \citealt{Casares2014} for a review).

Swift~J1727.8$-$1613 (hereafter J1727) is the latest BHT candidate to have been discovered (on August 24 2023; \citealt{Lipunov2023,Page2023}). Its brightness (reaching a peak outburst optical magnitude of $r \sim 12.7$; see \citealt{Wang2023,Baglio2023,Alabarta2023}) made it a target of observational campaigns at all wavelengths, which enabled the outburst evolution to be monitored. In the X-ray, J1727 followed the canonical behaviour for BH LMXBs (e.g. \citealt{Done2007,Belloni2011}), transitioning between  hard and soft X-ray states (see e.g. \citealt{Bollemeijer2023,Miller-Jones2023}). A number of studies dedicated to the X-ray event were soon published \citep{Chatterjee2024,Liu2024,Yu2024}, reporting properties such as the detection of quasi-periodic oscillations \citep{Zhu2024} and variable X-ray polarization \citep{Svoboda2024b,Podgorny2024b}. In radio, J1727 shows the largest resolved continuous radio jet found to date \citep{Wood2024}. Optical studies have revealed transient cold winds and enabled the identification of a pre-outburst optical counterpart, leading to early estimates of the orbital period ($P_{\rm orb}\sim 7.6\, {\rm h}$) and distance to the source ($d=2.7\pm 0.3\, {\rm kpc}$; \citealt{MataSanchez2024a}, MS24a hereafter).

The soft state outburst decay of J1727 continued in the X-ray until the low luminosity transition from the soft state to the hard one was reported (March 19 2024; \citealt{Russell2024,Podgorny2024a,Svoboda2024a}). Optical follow-up revealed a slowly dimming optical counterpart, reaching pre-outburst quiescence levels on June 1 2024 ($o\sim 19.6$, as is reported by the Asteroid Terrestrial-impact Last Alert System; \citealt{Tonry2018,Heinze2018,Smith2020,Shingles2021}), marking the end of a 10-month long outburst.

In this work, we present a time resolved spectroscopic campaign of J1727 in quiescence. We report on the detection of absorption features in the spectra consistent with a late-type star and measure its radial velocity curve. We obtain a precise determination of the orbital period and constrain the mass of the compact object. Additionally, we refine the distance and establish the peculiar velocity of the source, ultimately revealing the natal kick of the binary. In this work, uncertainties are cited at a $68\%$ confidence level unless otherwise specified.

\begin{table}
\caption{Journal of observations.}
\label{tab:obslog}
\begin{tabular}{ccccccccccc}
 Date  & Phase   &$$\#$$ &  Airmass & Seeing & $r$ \\
(dd/mm) & &   &  & (\arcsec) &   \\
\hline \\
 04/06 & $0.31-0.34$ & 2  &  1.41&1.1 &  $19.21 $\\
 05/06 & $0.67-0.70$ &  2  &  1.55&1.3 & $19.42$ \\
 07/06 & $0.09-0.11$  &  2  &  1.48 & 1.6 & $19.45 $ \\
 08/06 & $0.30-0.42$  &  6  &  1.46&1.3 & $19.43 $\\
 28/06 & $0.73-0.75$  &  2 & 1.42&0.7 & $19.60 $ \\
& $0.81-0.84$  & 2   & 1.51& 0.8 &  \\
 29/06 & $0.90-0.92$   & 2 & 1.42& 1.0 &  $19.59 $ \\
 30/06 & $0.99-1.01$   & 2 & 1.57& 1.0 &  $19.53 $ \\
&  $0.15-0.23$   & 4  & 1.43& 0.9 &  \\
 01/07 & $0.50-0.53$   & 2  & 1.59& 0.9 &  $19.75 $\\
 05/07 & $0.25-0.27$   & 2   & 1.42& 0.9&  $19.61 $\\
 06/07 & $0.58-0.60$  & 2  & 1.59 & 0.9&  $19.60 $ \\
 08/07 & $0.79-0.81$   & 2  & 1.44 & 1.4& $19.58 $ \\
 10/07 & $0.46-0.48$   & 2  & 1.63 & 1.3&  $19.77 $ \\
 12/07 & $0.78-0.80$   & 2  & 1.44& 0.9 & $19.62 $ \\
\end{tabular}
\tablefoot{The orbital phase is defined from the ephemerides derived in Sec \ref{sec:rv}. We report the number of exposures per epoch ($\#$), the airmass, the seeing, and the $r$-band magnitudes, measured from acquisition images. The latter have a typical uncertainty of $\pm 0.04$, dominated by the calibration of the companion star.}
\end{table}


\section{Observations} \label{sec:observations}

\subsection{Spectroscopy}

We observed J1727 just after the outburst demise (June 1), from June 4 to July 12. A total of 13 spectroscopic epochs containing 36 spectra of $900\,\rm{s}$ integration each (see Table \ref{tab:obslog}) were obtained with the \mbox{10.4-m} Gran Telescopio Canarias (GTC) at the Roque de los Muchachos Observatory (La Palma, Spain), equipped with the Optical System for Imaging and low-Intermediate-Resolution Integrated Spectroscopy (OSIRIS, \citealt{Cepa2000}). We employed a $0.8\arcsec$ slit width, $2\times 2$ binning, and grism R2500R, resulting in a wavelength coverage of $5560 - 7670\, {\rm \AA}$. We reduced the dataset using \textsc{pyraf} and \textsc{molly} tasks, correcting from bias, flats, calibrating in wavelength with arc lamps, and optimally extracting the spectra. We found sub-pixel drifts on the position of the [\ion{O}{i}]-6300 sky emission line, which were employed to correct for flexure effects on the wavelength calibration. We corrected the spectra from the Earth movement using \textsc{molly} tasks, which sets all derived velocities in this work in the heliocentric reference frame. We measured the seeing from the spatial profile of the bidimensional spectra, performing a simultaneous Gaussian fit to the target and a near field star placed in the slit ($2.7''$ to the north-east; see Fig. 4 in M24a). We limited the calculation to the spectral range of interest ($5900-6500\, \AA$), allowing us to conclude that our data is slit-limited (Table \ref{tab:obslog}). The spectral dispersion is $1.02\, \AA / {\rm pix}$, while the average spectral resolution is measured from arc lines to be $R=\lambda /\Delta \lambda =2900$. We normalised the spectra through a fit to the continuum using a seventh-order polynomial. 

Stellar templates of spectral types F9V-K5V and solar metallicity ($\pm 0.1\ \rm{dex}$) were observed during previous campaigns with grism R2500R, slit $0.6''$ (see Table \ref{tab:templ}). They were all reduced and normalised following the same steps as for our target. Additionally, they were corrected from their observed radial velocities, as is reported in \citet{Gaia2023}, setting them all in a null-velocity frame. Finally, a Gaussian fit to the H$\alpha$ absorption line was performed in the templates to detect and correct for any remaining flexure offset.

\begin{table*}
\caption{Radial velocity curve parameters (uncertainties at $1\sigma$).}
\label{tab:templ}
\begin{tabular}{lcccccccccc}
 Template  & $T_{\rm eff}$   & SpecType & $K_2$ & $\gamma$ & $t_0$  & $P_{\rm orb}$ & $\chi_{\rm red}^2$ &  \\
 & (K) &   & $\rm (km\, s^{-1})$  & $\rm (km\, s^{-1})$ &  (d) & (d) &  \\
\hline \\
HD 19373   & 5991 & F9/G0~V  & $389\pm 4$ & $-185\pm 3$ &  $-0.4684\pm 0.0020$ &  $0.45016  \pm 0.00003$ & 0.97 \\
HD 186408  & 5815 & G2~V  & $390\pm 4$ & $-180\pm 3$ &  $-0.4688\pm 0.0021$ &  $0.45016  \pm 0.00004$ & 1.00 \\
HD 186427  & 5762 & G2~V  & $388\pm 4$ & $-182\pm 3$ &  $-0.4682\pm 0.0023$ & $0.45016  \pm 0.00004$ & 1.06  \\ 
HD 235299  & 5471 &  G8~V & $392\pm 4$ & $-184\pm 3$ &  $-0.4696\pm 0.0021$ &  $0.45018  \pm 0.00004$ & 1.08 \\
HD 185144  & 5260 & K0~V   & $390\pm 4$ & $-184\pm 3$ &  $-0.4694\pm 0.0020$ & $0.45018  \pm 0.00003$ & 1.12 \\
HD 219134  & 4717 & K3/4~V  & $390\pm 3$ & $-179\pm 3$ &  $-0.4681\pm 0.0019$ &  $0.45015  \pm 0.00003$ & 0.89 \\
HD 36003  & 4465 & K5~V & $390\pm 3$ & $-176\pm 3$ &  $-0.4685\pm 0.0020$ & $0.45016  \pm 0.00003$ & 0.98 \\
\end{tabular}
\tablefoot{For the template stars we adopt the $T_{\rm eff}$ reported in MILES spectral library \citep{MILES2006}, except for HD 235299 \citep{Soubiran2016}. The spectral type is derived from this $T_{\rm eff}$ and tabulated values in \citet{Pecaut2013}. The best fitting parameters for the sinusoidal function described in Sect. \ref{sec:rv} are also reported, being $t_0=T_0 \rm{(HJD)}-2460465.0$, and defining $\chi_{\rm red}^2=\chi^2 / (32\, \rm{d.o.f.})$. }
\end{table*}

\subsection{Photometry}

We analysed the $r$-band acquisition images obtained prior to each set of spectra. Point spread function (PSF) photometry was performed with the \textsc{DAOPHOT} package \citep{Stetson1987} in \textsc{IRAF}. A Moffat distribution model was constructed for five selected field stars after removing their neighbours. We performed differential photometry on the target relative to a comparison star (PSO J261.9304$-$16.2053, $r = 19.38 \pm 0.01$; as reported in Pan-STARRS1 Survey, \citealt{PanSTARRS2016}). The resulting photometry is shown in Table \ref{tab:obslog}, and yields a mean magnitude and root-mean-square variability of $r = 19.55 \pm 0.14$ for J1727 during our observations.

\section{Analysis and results} \label{sec:analysis}

\begin{figure}
\centering
\includegraphics[keepaspectratio, trim=2.0cm 0cm 2.5cm 2cm, clip=true, width=0.5\textwidth]{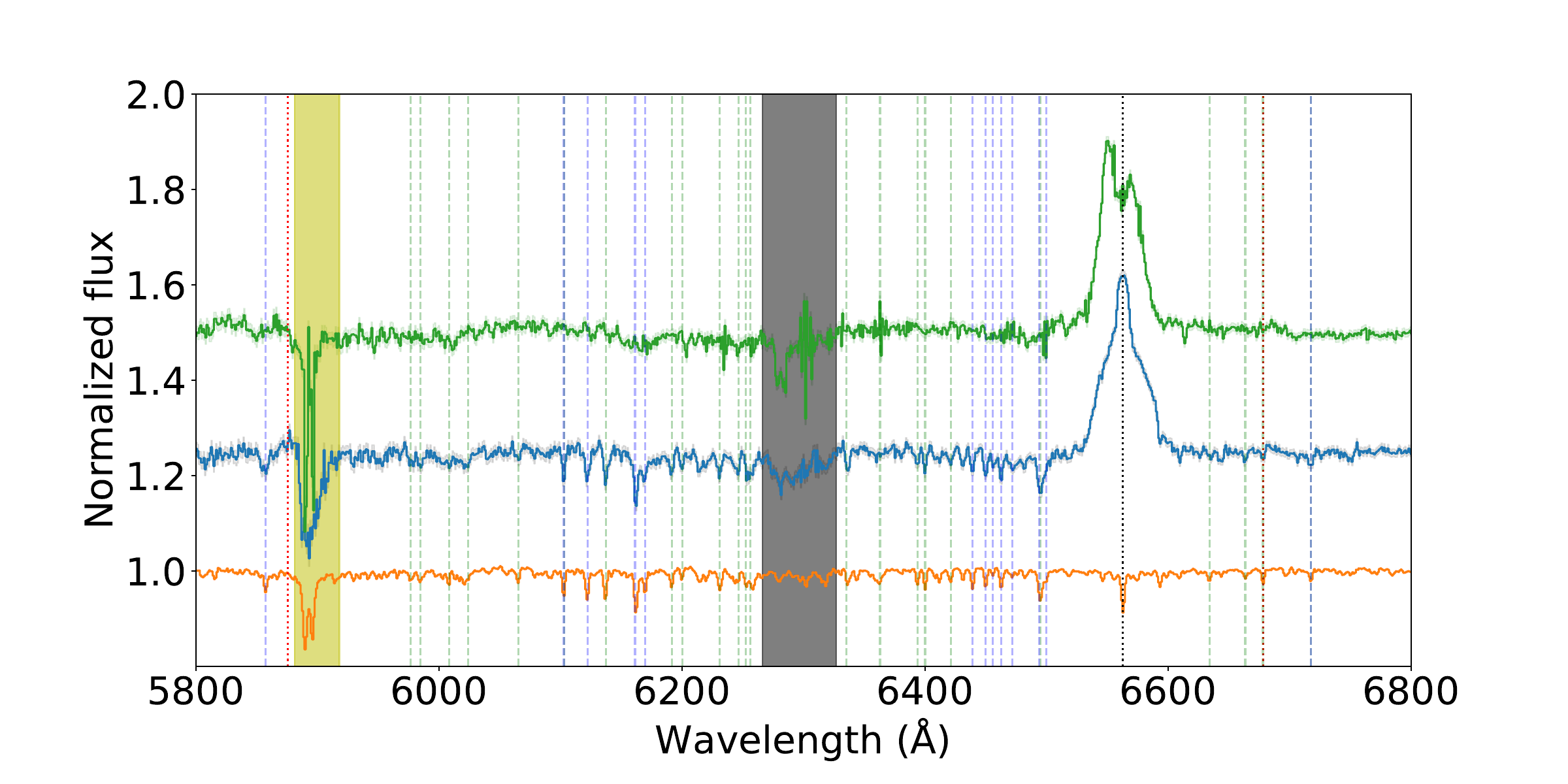}
\caption{Normalised, averaged spectrum of J1727 in the observer rest frame (top, solid green line), as well as in the companion star rest frame (middle, solid blue line). A template K3/4V star (HD219134) is shown in orange after applying the veiling factor derived in this work ($X_{\rm veil}=0.74$). They are offset vertically by 0.25 for visualisation purposes. Absorption features identified with known atomic transitions in this wavelength range are marked as dashed lines in blue for \ion{Ca}{i}, and green for \ion{Fe}{i} and \ion{Fe}{ii}. Disc emission lines are marked with dotted lines, black for H  and red for \ion{He}{i}. Telluric bands and interstellar absorption are marked with shaded grey and yellow regions, respectively. Note that absorption features associated with the companion star are revealed as narrow features only in the companion rest frame.}
\label{fig:avenorm}
\end{figure}

The defining feature of the spectra is a double-peaked, asymmetric H$\alpha$ emission line. No other remarkable features are seen in the normalised averaged spectrum in the observer's rest frame (see Fig. \ref{fig:avenorm}). In what follows, we shall analyse the individual and averaged spectra in the rest frame of the companion star.

\subsection{Radial velocity curve} \label{sec:rv}

\begin{figure}
\centering
\includegraphics[keepaspectratio, trim=2.3cm 0cm 2.5cm 2cm, clip=true, width=0.5\textwidth]{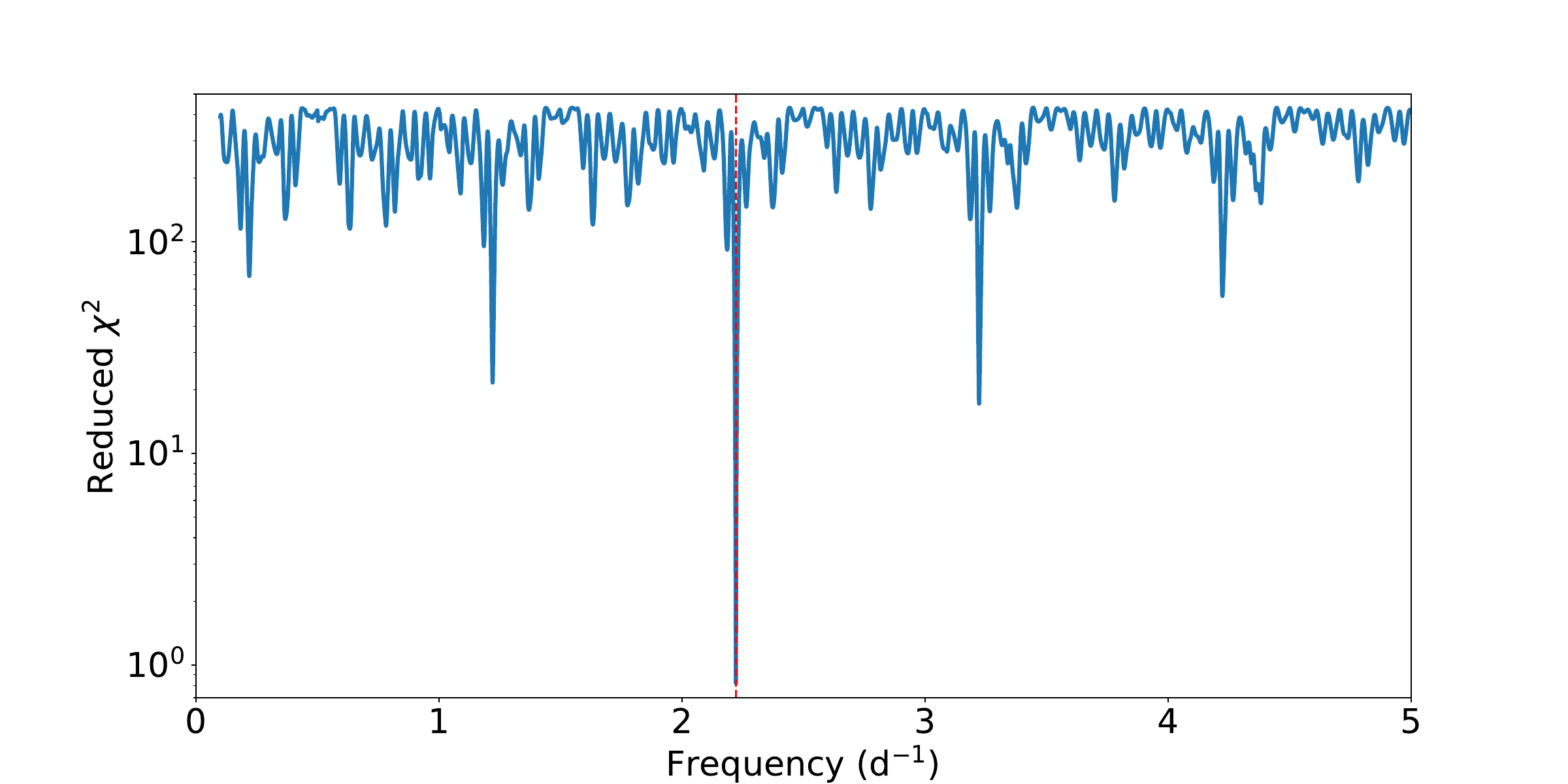}
\caption{Reduced $\chi^2$ periodogram of the radial-velocity curve generated from the cross-correlation with HD 219134 (K3/4V). The best-fitting period is marked with a dashed red line. Note the logarithmic scale in the y axis.}
\label{fig:periodogram}
\end{figure}

\begin{figure}
\centering
\includegraphics[keepaspectratio, trim=1.6cm 0cm 2.5cm 2cm, clip=true, width=0.5\textwidth]{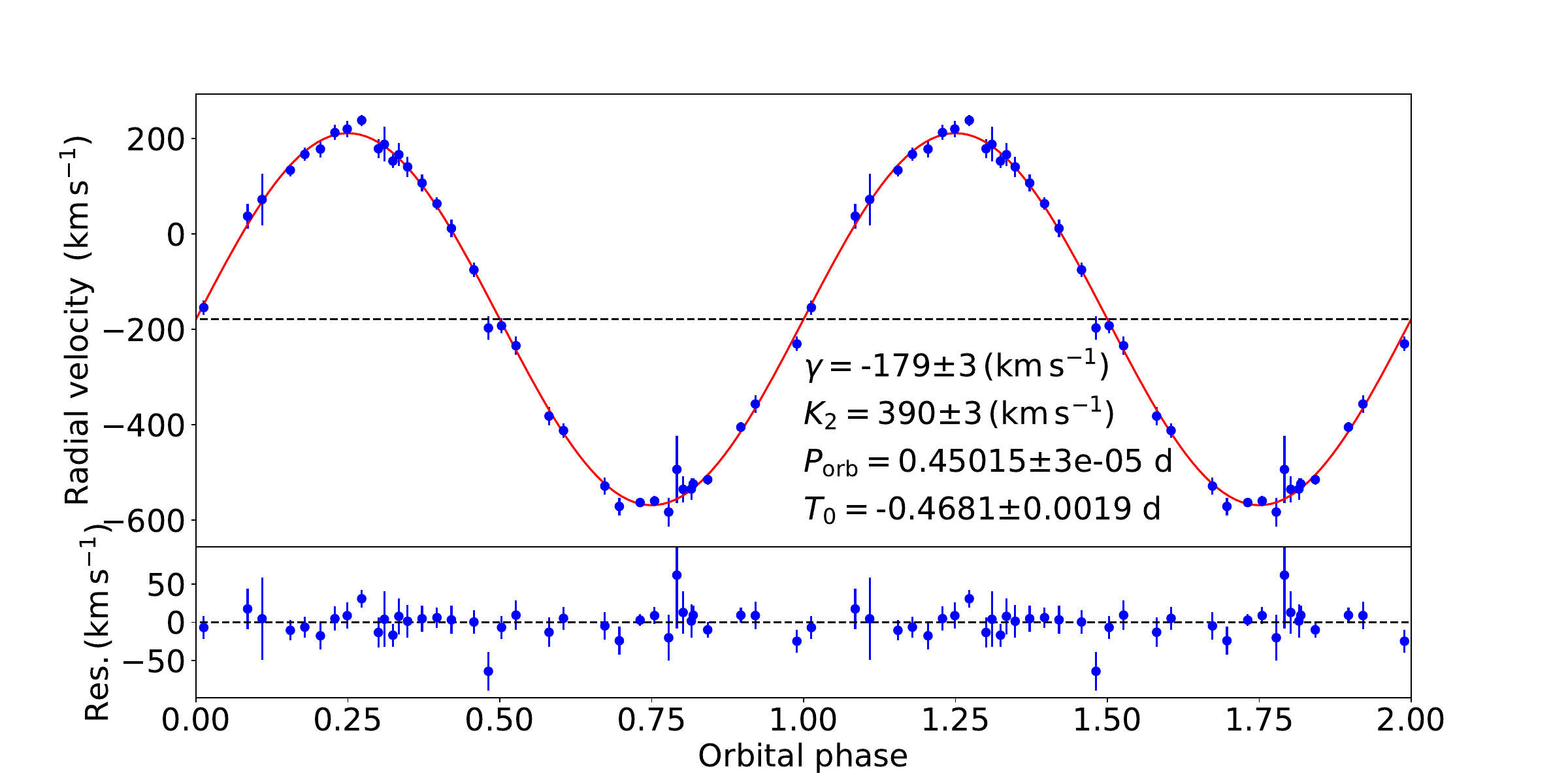}
\caption{Radial velocity curve of the companion star in J1727 measured through cross-correlation with the template star HD219134 (K3/4V), and folded to the best-fit ephemerides. We repeated two orbital phases for visualisation purposes. The top panel shows the individual radial velocities (blue dots), as well as the best fit sinusoid (solid red line). The best-fitting parameters are also reported in this panel. The bottom panel shows the residuals from the best fit.}
\label{fig:rv}
\end{figure}

We searched for the companion star via cross-correlation of the J1727 individual spectra with the K3/4 V template HD219134 in the wavelength range $ 5920-6820\,\AA$ (masking out both the telluric band at $6300\,\AA$ and H$\alpha$). We derived the cross-correlation functions (CCFs) using the cross-corrRV \textsc{python} routine from \textsc{PyAstronomy}\footnote{https://github.com/sczesla/PyAstronomy}. The CCFs peaks were fitted with a single Gaussian model to determine their centroids, while uncertainties on the measured radial velocities were estimated through a Monte Carlo analysis, as is described in \citet{MataSanchez2021}. This yielded the radial velocity curve of the companion star. 

We searched for periodicities in the radial velocity dataset using a $\chi^2$ periodogram, inspecting the frequency range of $0.1-5\, \rm{d^{-1}}$ in $10^{-5} \, \rm{d^{-1}}$ frequency steps (Fig. \ref{fig:periodogram}).
We identify a prominent peak at $\sim 0.45\,\rm{d}$. The secondary peaks correspond to aliases arising from a combination of the daily repetition of the observing window with the observing cadence of $\sim 15\, \rm{min}$. Indeed, attempting to perform a sinusoidal fit with a starting period value from any of the secondary peaks leads to poorly constrained fits ($\chi_{\rm red}^2 > 20$). Only the strongest peak produces a radial velocity curve with a sinusoidal pattern, which we model as:
$$v(t)=\gamma + K_2\sin{((t-T_0)/P_{\rm{orb}})},$$where $P_{\rm{orb}}$ is the binary orbital period, $K_2$ the radial velocity semi-amplitude of the companion star, $\gamma$ the systemic velocity, $t$ the observation heliocentric Julian date (HJD), and $T_0$(HJD) the reference time at the inferior conjunction of the companion. Figure \ref{fig:rv} shows the radial velocity curve folded on the obtained ephemeris.

We repeated the above process to compare J1727 with each template of Table \ref{tab:templ}, covering spectral types F9V to K5V. The best-fitting parameters obtained are fully consistent within uncertainties, HD 219134 (K3/4V) being the best-fitting template. The largest spread is shown by $\gamma$ (still consistent within $1.5\sigma$), which we associate with uncertainties in the correction of their radial velocities. Adding the normal distributions for the parameters obtained for each template (see Table \ref{tab:templ}) leads to the following $1\sigma$ solution:

$$\gamma= -181 \pm 4 \,\rm{km\, s^{-1}}  \qquad  \textit{K}_2= 390 \pm 4 \,\rm{km\, s^{-1}}$$
$$P_{\rm{orb}}= 0.45016 \pm 0.00004\, \rm{d}$$ 
$$\textit{T}_0 \rm{(HJD)}= 2460464.5313 \pm 0.0021\, \rm{d}.$$

\subsection{Companion star spectral type} \label{sec:sptype}

The orbital solution allows us to correct the individual spectra from the orbital motion, and produce an average normalised spectrum in the companion rest frame (Fig. \ref{fig:avenorm}, middle spectrum). We employed it for its spectral type classification, considering the same wavelength range as for the cross-correlation and following the approach described in \citet{MataSanchez2021}, based on the \textsc{emcee} Markov chain Monte Carlo sampler \citep{Foreman-Mackey2013}. This method consists of comparing the average spectra in the companion rest frame with a synthetic grid of templates constructed from the sample in \citet{Coelho2014}. The templates were degraded to the instrumental resolution of our observations ($R=2900$), and we restricted our analysis to solar metallicity models. The synthetic templates are defined by the following parameters: the effective temperature ($T_{\rm eff}$), surface gravity ($\log{g}$), rotational broadening ($v_{\rm \sin i}$), and velocity shift ($v_{\rm shift}$ to account for potential offsets given the dispersion in $\gamma$ with the spectral type of Table \ref{tab:templ}), and veiling factor ($X_{\rm veil}$), defined as the ratio of fluxes of non-stellar origin to the total emitted light at the observed range of wavelengths (covering the $r$ band). 

We set the same uniform priors on the parameters as are described in \citet{MataSanchez2021}. These put loose but physical constraints on the parameter space, and result in the corner plot in Fig. \ref{fig:corner}. We report the 16, 50, and 84$\%$ confidence levels of the marginalised distributions for the inspected parameters (corresponding to $1\sigma$ uncertainties on a Gaussian distribution). For those parameters for which only a lower or upper limit is provided, we report the 16 or 84$\%$ confidence levels, respectively:

$$T_{\rm eff}=4570^{+90}_{-100}\, {\rm K} \qquad X_{\rm veil}= 0.74\pm 0.02$$
$$v_{\rm \sin i}<52\, \rm{km\, s^{-1}} \qquad v_{\rm shift}=-4\pm 4\, \rm{km\, s^{-1}} \qquad \log{g}> 4.45.$$

\noindent $T_{\rm eff}$ is well constrained and corresponds to a spectral type of K4$\pm$1 V for the companion star \citep{Pecaut2013}, in line with the best-fitting template used in the cross-correlation analysis. The relatively high veiling factor ($X_{\rm veil}$) implies a large contribution from the accretion disc. In this regard, the optical counterpart was 0.4 magnitudes brighter than in the pre-outburst imaging (MS24a). We cannot confidently measure the rotational broadening, which suggests a relatively low value (e.g. comparable or lower than our spectral resolution). The analysis also favours a lower limit of $\log{g}$ typical of dwarf stars. Finally, the $v_{\rm shift}$ is consistent with zero, indicating that $\gamma$ was properly determined in the previous section.

\section{Discussion} \label{sec:discussion}

The analysis of the J1727 spectra during early quiescence enabled the measurement of key parameters of the system, which are discussed below.

\subsection{From $P_{\rm orb}$ to the black hole mass}

We directly measured $P_{\rm orb}= 10.8038\pm 0.0010\, {\rm h}$, longer than the value predicted in MS24a ($P_{\rm orb}\sim 7.6\, {\rm h}$), using an empirical relation between $P_{\rm orb}$ and the outburst optical amplitude (see \citealt{Shahbaz1998}; also \citealt{Lopez2019} for a discussion). Together with $K_2$, it allows us to derive the mass function, which imposes a conservative lower limit on the compact object mass:
$$f(M_{1})= \dfrac{M_1 \, \sin{^3 i}}{(1+q)^2} =\dfrac{K_2^3 P_{\rm orb}}{2\pi G}=2.77\pm 0.09 \, M_\odot; \quad q=\dfrac{M_2}{M_1},$$
where $M_1$ and $M_2$ are the compact object and the companion star masses, respectively, $q$ is the mass ratio, and $i$ is the orbital inclination.

A conservative theoretical upper limit of $\lesssim 3\, M_\odot$ \citep{Kalogera1996} is often employed to discriminate NSs from BHs. However, we find that $f(M_{1})=2.77\pm 0.09 \, M_\odot$, which is a strict lower limit on the compact object mass assuming the limiting case of $q\sim 0$ and $i=90\,\deg$. Sensible constraints on both parameters are discussed below in order to fully confirm the BH nature of the compact object.

The determination of a precise $q$ value is traditionally achieved by measuring the rotational broadening of the companion star absorption lines (see e.g. \citealt{Torres2020}). However, this requires templates taken with the same instrumental set-up as the target observations, which we lack. We can set constraints on $q$ using the spectral type derived from our analysis (K4$\pm$1 V). Previous studies have revealed that most LMXB companions are not main-sequence stars, but instead slightly evolved (see \citealt{Rappaport1983,Podsiadlowski2003}). As a result, the mass associated with the derived spectral type is better understood as an upper limit on the true companion star mass, which for J1727 results in $M_2<0.78\,M_\odot$. A direct comparison can be drawn with GRS 1124$-$684, a BH LMXB with a similar $P_{\rm orb}\sim 10.4\, {\rm h}$ \citep{GonzalezHernandez2017}, donor spectral type K3-5 V \citep{Orosz1996,Casares1997}, and $q=0.079\pm 0.007$ \citep{Wu2015}. A similar mass ratio in J1727 would imply $M_1>3.22\pm 0.10\, M_\odot$, above the NS threshold. 

Arguably the most critical parameter to fully solve the dynamical mass of LMXBs is $i$. The BH nature of the accretor in J1727 is independently confirmed if $i<74\, \deg$ ($q\sim 0$); that is, as long as an edge-on scenario is rejected. For non-eclipsing LMXBs, direct measurement of $i$ is traditionally done through photometric light curve modelling of the ellipsoidal modulation produced by the distorted companion star (e.g. \citealt{Casares2014}). We do not detect this periodic variability in our photometry. We rather propose that the variability of the light curve ($\sigma_r=0.14$) is instead dominated by flickering of the accretion disc (see e.g. \citealt{Zurita2003}). The equivalent width of the H$\alpha$ profile across our sample varies between $EW=10-21\, {\rm \AA}$, with the largest values observed at the faintest epochs, as was expected from variability dominated by the accretion disc fading in brightness towards its quiescence. No modulation in the $EW$ akin to that observed in the BHT MAXI J1820+070 \citep{Torres2019} was found at the $P_{\rm orb}$, arguing against a high-inclination configuration.

X-ray observations allow us to further explore this elusive parameter. X-ray dips, a feature observed in LMXBs with inclinations $>60\, \deg$ \citep{Frank1987}, have not been reported for J1727, despite the intense monitoring enabled by its bright outburst event. Finally, radio observations of the source revealed the largest resolved continuous jet ever discovered in a LMXB \citep{Wood2024}, and imposed an upper limit on the jet inclination of $<74\, \deg$. Assuming that the jet is launched perpendicular to the orbital plane, it would serve as proxy for $i$. Setting this constraint alone implies a compact object mass of $M_1> 3.12\pm 0.10 \, M_\odot$, placing it above the most conservative mass threshold at $1\sigma$. We therefore conclude that all the available evidence supports the BH nature of J1727.

\subsection{The distance and the peculiar velocity of J1727}

MS24a proposed a distance of $2.7\pm 0.3\, {\rm kpc}$ from a number of arguments based on both optical and X-ray observations. Except for the method that directly provides the distance based on the interstellar \ion{Ca}{ii} doublet EW ratio ($d=3.2\pm 0.6\, \rm{kpc}$), the rest rely on determining the reddening and $P_{\rm orb}$, with the latter utilised to derive the absolute magnitude of the optical counterpart ($M_{\rm r}$, \citealt{Casares2018a}). We updated the results from MS24a using our spectroscopic measurement of $P_{\rm orb}$, which makes the system brighter ($M_{\rm r}=5.92\pm 0.11$) and consequently increases the final estimated distance. Estimates for the colour excess of $E(B-V)=0.9\pm 0.3$ and $0.8\pm 0.3$ were found from interstellar absorption features in the outburst optical spectra, while $E(B-V)=0.47\pm 0.13$ were found from modelling of X-ray data, all of which are reported in MS24a. Dust maps from the Pan-STARRS1 Survey \citep{Green2019} reveal a moderate distance dependence, $E(B-V)=0.23\pm 0.04$ for $d=1-4\,{\rm kpc}$ and $E(B-V)=0.33\pm 0.03$ for larger distances. This is lower than any of the results from the aforementioned dedicated methods, which suggests that the extinction towards J1727 is dominated by an unresolved interstellar structure in the dust maps. Taking into account these estimates for the colour excess, we updated the distance towards J1727 to $d=2.6\pm 0.9\,{\rm kpc}$, $2.9\pm 1.0\,{\rm kpc}$, and $4.0\pm 0.6\,{\rm kpc}$, respectively. Moreover, we can use the results from Sect. \ref{sec:analysis} to calculate the spectroscopic distance assuming a dwarf companion of spectral type K4V. Tabulated absolute magnitudes exist \citep{Pecaut2013}, which we transformed following \citet{Jordi2006} to obtain an absolute magnitude in the ${\rm r}$-band of $6.6\pm 0.5$. To properly derive the distance from this value, we need the companion star apparent magnitude. This is $r=21.13\pm 0.05$, obtained from the observed magnitude in our acquisition images ($r=19.55\pm 0.14$) and the veiling factor derived from the spectroscopy ($X_{\rm veil}=0.74\pm 0.02$). Correcting from reddening with the same set of colour excess estimates introduced above, it results in $d=3.2\pm 1.3\,{\rm kpc}$, $3.5\pm 1.5\,{\rm kpc}$, and $4.8\pm 1.4\,{\rm kpc}$, respectively.

The spread in the resulting $d$ from each of the above methods highlights how complicated it is to measure a precise distance in the absence of parallax. Assuming all of the aforementioned methods are equally reliable, we calculated the conditional probability through the multiplication of the individual distributions. We obtained a final distance, fully coincident with the weighted average and weighted standard deviation, of $d=3.4\pm 0.3\,{\rm kpc}$ (i.e. $|z|=0.61\pm 0.06\, \rm{kpc}$ below the Galactic plane). We employed this result hereafter in our calculations; it corresponds to a range of $d=2.5 - 4.3\,{\rm kpc}$ at the $3\sigma$ confidence level. Using our favoured $d$, the soft-to-hard state transition ($\sim 0.1-1.2\,\rm{cts\,s^{-1}}$,
\citealt{Podgorny2024b}) occurs at $0.01-0.02\, L_{\rm Edd}$ ($L_{\rm Edd}$ being the Eddington luminosity of a canonical BH)\footnote{We calculate $L_{\rm Edd}$ for a prototypical BH ($M_1=8\, M_\odot$), and convert it to flux units using $d=3.4\pm 0.3\,{\rm kpc}$. We also apply a bolometric correction of $2.9$ (with $25\%$ uncertainty, see \citealt{intZand2007}), and finally convert to $\rm{cts\,s^{-1}}$ for the MAXI GSC instrument in the 2-20 keV band using the WebPIMMS tool [\url{https://heasarc.gsfc.nasa.gov/cgi-bin/Tools/w3pimms/w3pimms.pl}].}, within the expected range (see \citealt{Maccarone2003,Dunn2010}).

An astrometric solution for the optical counterpart to J1727 is reported in \textit{Gaia} Data Release 3 \citep{Gaia2016,Gaia2023}. While the proposed parallax is not particularly constraining, there is a proper motion reported for the system of $\Delta\alpha =-0.04\pm 0.60\, \rm{mas}\,\rm{yr}^{-1}$ and $\Delta\delta =-4.92\pm 0.43\,\rm{mas}\,\rm{yr}^{-1}$. Combined with the distance ($d=3.4\pm 0.3\, \rm{kpc}$) and systemic velocity ($\gamma=-181\pm 4\, \rm{km\, s^{-1}}$) measured in the present work, we derived the space velocity in the local standard of rest $(U,V,W)=(-148\pm 5,-88\pm 10,-71\pm 10)\,\rm{ km\,s^{-1}}$. The latter is based on \citet{Johnson1987} transformations and assumes a solar space velocity in the same rest frame of $(U_\odot,V_\odot,W_\odot)=(9,12,7)\,\rm{ km\,s^{-1}}$ \citep{Mihalas1981}, resulting in a space velocity of $v_{\rm space}=194\pm 5\,\rm{ km\,s^{-1}}$ for J1727. We note that the dominant contribution to $v_{\rm space}$ and its uncertainty is the radial velocity component ($\gamma$). We can transform this result to a peculiar velocity referred to the J1727 Galactic neighbourhood, assuming the galactocentric potential \textsc{MWPotential2014} from the \textsc{galpy} package \citep{Bovy2015}, obtaining $(U_{\rm{pec}} ,V_{\rm{pec}},W_{\rm{pec}})=(-174\pm 4,-75\pm 10,-66\pm 10)\,\rm{ km\,s^{-1}}$. The large resulting $v_{\rm pec}=201\pm 7\,\rm{ km\,s^{-1}}$ suggests that a kick was imparted to the system at birth. Following \citet{Atri2019}\footnote{We employed the public script available at the github link
https://github.com/pikkyatri/BHnatalkicks.} to integrate the Galactocentric orbit 10 Gyr back in time, we obtain a distribution of potential natal kick velocities, with 5, 50, and 95$\%$ percentiles of $v_{\rm kick}=210^{+40}_{-50}\,\rm{ km\,s^{-1}}$. This is consistent with the upper end of natal kicks for the BH population presented in the aforementioned work, and further supports supernova models over direct collapse for the birth of the BH in J1727.

\section{Conclusions}

We present a spectroscopic study of the recently discovered BHT
candidate J1727 in early quiescence. Phase-resolved spectroscopy reveals stellar features from a veiled ($X_{\rm veil}=0.74\pm 0.02$) K4V ($T_{\rm eff}=4570^{+90}_{-100}\, {\rm K}$) companion star, which allows us to perform a dynamical study. We measure $P_{\rm{orb}}= 0.45016 \pm 0.00004\, \rm{d}$ and $K_2= 390 \pm 4 \,\rm{km\, s^{-1}}$,  which results in a mass function of $f(M_{1})=2.77\pm 0.09 \, M_\odot$. We present arguments against a fully edge-on system geometry, which allow us to set a robust lower limit on the compact object mass of $M_1> 3.12\pm 0.10 \, M_\odot$, confirming the BH nature of the compact object. We further refine the distance to the system with our updated binary parameters to be $3.4\pm 0.3\, {\rm kpc}$. Combined with the systemic velocity and previous reports on the proper motion, we obtain a natal kick for the BH of $v_{\rm kick}=210^{+40}_{-50}\,\rm{ km\,s^{-1}}$ (at 5, 50, and 95$\%$ percentiles). 

\begin{acknowledgements}

DMS, MAPT and TMD  acknowledge support by the Spanish Ministry of Science via the Plan de Generacion de conocimiento PID2021-124879NB-I00, and JC via PID2022-143331NB-100.
MAP acknowledges support through the Ramón y Cajal grant RYC2022-035388-I, funded by MCIU/AEI/10.13039/501100011033 and FSE+. We are thankful to the GTC staff for performing the observations. This work has made use of data from the European Space Agency (ESA) mission {\it Gaia} (\url{https://www.cosmos.esa.int/gaia}), processed by the {\it Gaia} Data Processing and Analysis Consortium (DPAC, \url{https://www.cosmos.esa.int/web/gaia/dpac/consortium}). Funding for the DPAC has been provided by national institutions, in particular the institutions participating in the {\it Gaia} Multilateral Agreement. We are deeply grateful to Tom Marsh for developing the \textsc{molly} software, one of his many contributions to advancing the field of compact objects. \textsc{pyraf} is the python implementation of \textsc{iraf} maintained by the community. The Pan-STARRS1 Surveys (PS1) and the PS1 public science archive have been made possible through contributions by the Institute for Astronomy, the University of Hawaii, the Pan-STARRS Project Office, the Max-Planck Society and its participating institutes, the Max Planck Institute for Astronomy, Heidelberg and the Max Planck Institute for Extraterrestrial Physics, Garching, The Johns Hopkins University, Durham University, the University of Edinburgh, the Queen’s University Belfast, the Harvard-Smithsonian Center for Astrophysics, the Las Cumbres Observatory Global Telescope Network Incorporated, the National Central University of Taiwan, the Space Telescope Science Institute, the National Aeronautics and Space Administration under Grant No. NNX08AR22G issued through the Planetary Science Division of the NASA Science Mission Directorate, the National Science Foundation Grant No. AST1238877, the University of Maryland, Eotvos Lorand University (ELTE), the Los Alamos National Laboratory, and the Gordon and Betty Moore Foundation. This work has made use of data from the Asteroid Terrestrial-impact Last Alert System (ATLAS) project. The Asteroid Terrestrial-impact Last Alert System (ATLAS) project is primarily funded to search for near earth asteroids through NASA grants NN12AR55G, 80NSSC18K0284, and 80NSSC18K1575; byproducts of the NEO search include images and catalogues from the survey area. This work was partially funded by Kepler/K2 grant J1944/80NSSC19K0112 and HST GO-15889, and STFC grants ST/T000198/1 and ST/S006109/1. The ATLAS science products have been made possible through the contributions of the University of Hawaii Institute for Astronomy, the Queen’s University Belfast, the Space Telescope Science Institute, the South African Astronomical Observatory, and The Millennium Institute of Astrophysics (MAS), Chile. We thank the referee for their useful comments that helped improved the manuscript.
\end{acknowledgements}

%
%

\bibliographystyle{aa} 
\bibliography{bibliography} 

\begin{appendix} 
\section{Additional figure}

\begin{figure*}
\centering
\includegraphics[keepaspectratio, trim=0cm 0cm 0cm 0cm, clip=true, width=1.0\textwidth]{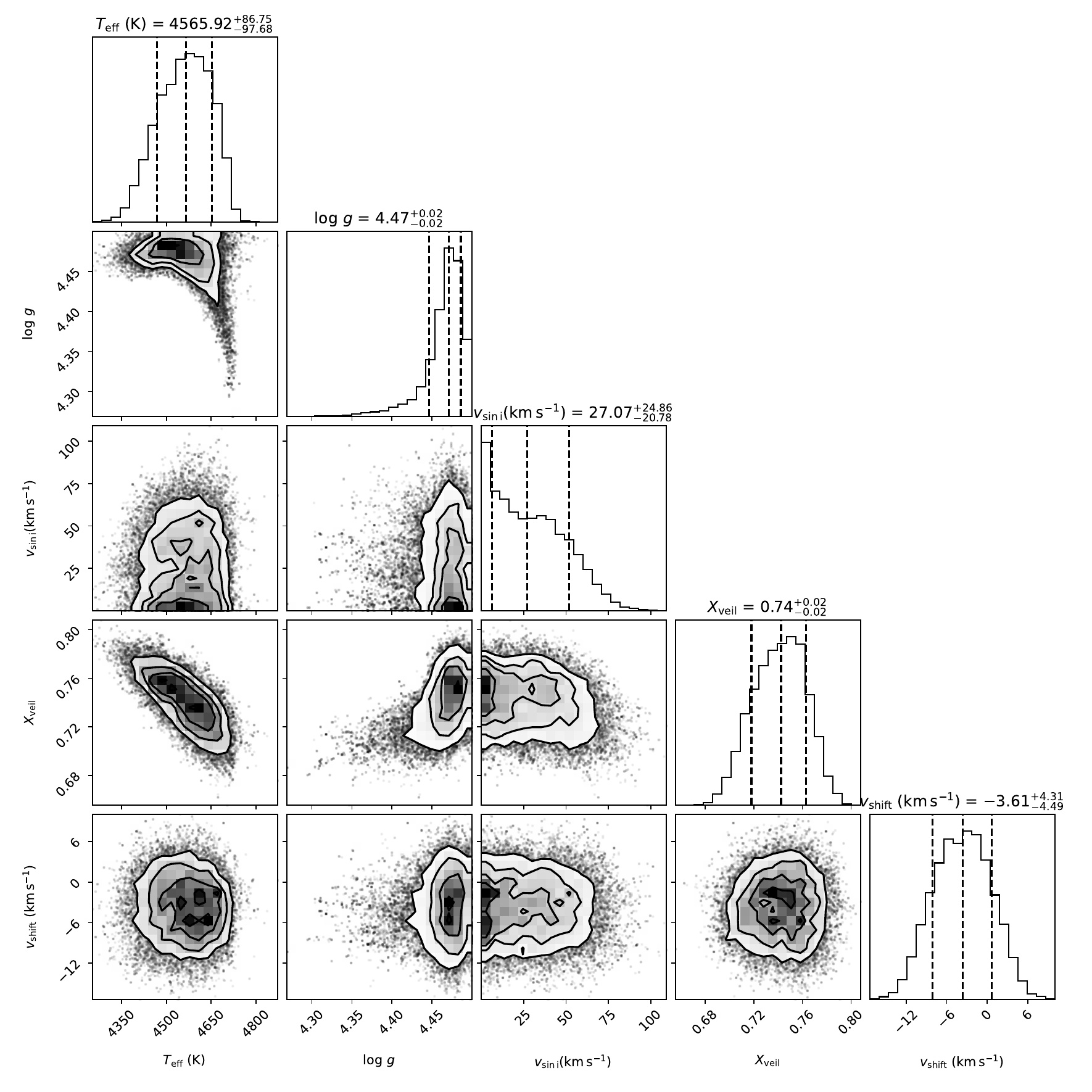}
\caption{Corner plot for the free parameters of the EMCEE spectral classification. Contours in the 2D plots correspond to the 0.5$\sigma$, 1$\sigma$, 1.5$\sigma,$ and 2.5$\sigma$ levels (respectively containing 11.8, 39.3, 67.5, and 86.5 $\%$ of the volume), while in the marginalised 1D distributions the 0.16, 0.50, and 0.84 quantiles are marked with dashed lines.}
\label{fig:corner}
\end{figure*}

\end{appendix}

\end{document}